\documentclass{PoS}
\let\ifpdf\relax
\usepackage[authoryear,square]{natbib}
\bibpunct{(}{)}{;}{a}{}{,}
\usepackage{subfig}

\title{Synergy between the Large Synoptic Survey Telescope and the Square Kilometre Array}

\ShortTitle{LSST-SKA Synergy}

\author{\speaker{David Bacon}$^1$,
Sarah Bridle$^2$,	
Filipe B. Abdalla$^{3,4}$,
Michael Brown$^2$,
Philip Bull$^5$,
Stefano Camera$^{2,6}$,
Rob Fender$^7$,
Keith Grainge$^2$,
\v{Z}eljko Ivezi\'{c}$^8$,
Matt Jarvis$^{7,9}$,
Neal Jackson$^2$,
Donnacha Kirk$^3$,
Bob Mann$^{10}$,
Jason McEwen$^{11}$,
John McKean$^{12}$,
Jeffrey A. Newman$^{13}$,
Alvise Raccanelli$^{14,15,16}$,
Martin Sahl\'en$^7$,
Mario Santos$^9$,
Anthony Tyson$^{17}$,
Gong-Bo Zhao$^{18,1}$
\\	
$^1$University of Portsmouth, UK;
$^2$University of Manchester, UK;
$^3$University College London, UK;
$^4$Rhodes University, South Africa;
$^5$University of Oslo, Norway;
$^6$University of Lisbon, Portugal;
$^7$University of Oxford, UK;
$^8$University of Washington, USA;
$^9$University of the Western Cape, South Africa;
$^{10}$University of Edinburgh, UK;
$^{11}$Mullard Space Science Laboratory, University College London, UK;
$^{12}$ASTRON, The Netherlands;
$^{13}$University of Pittsburgh, USA;
$^{14}$Johns Hopkins University, USA;
$^{15}$Jet Propulsion Laboratory, Caltech, USA;
$^{16}$California Institute of Technology, USA;
$^{17}$University of California, Davis, USA;
$^{18}$National Astronomy Observatories, Beijing, China.
\\
E-mail: \email{david.bacon at port.ac.uk}}

\abstract{We provide an overview of the science benefits of combining
information from the Square Kilometre Array (SKA) and the Large Synoptic
Survey Telescope (LSST). We first summarise the capabilities and timeline
of the LSST and overview its science goals. We then discuss the science
questions in common between the two projects, and how they can be best
addressed by combining the data from both telescopes. We describe how weak
gravitational lensing and galaxy clustering studies with LSST and SKA can
provide improved constraints on the causes of the cosmological
acceleration. We summarise the benefits to galaxy evolution studies of
combining deep optical multi-band imaging with radio observations. Finally,
we discuss the excellent match between one of the most unique features of
the LSST, its temporal cadence in the optical waveband, and the time
resolution of the SKA.  }

\FullConference{
Advancing Astrophysics with the Square Kilometre Array\\
June 8-13, 2014\\
Giardini Naxos, Italy}

\begin{document}

\section{The Large Synoptic Survey Telescope}

\subsection{The LSST system}

\noindent
The LSST system is designed to achieve multiple goals in four main science themes: (i) taking an 
inventory of the Solar System; (ii) mapping the Milky Way; (iii) exploring the transient optical 
sky; and (iv) probing dark energy and dark matter. These are just four of the many areas in 
which LSST will have enormous impact, but they span the space of technical challenges 
in the design of the system and survey, and have been used to focus the science requirements.
The LSST will be a large, wide-field ground-based telescope, camera and data management system
designed to obtain multi-band images over a substantial fraction of the sky every few nights.
The observatory site will be located on Cerro Pach\'on in northern Chile (near the Gemini South 
and SOAR telescopes), with the first light expected in 2021 and the first public data releases
mid-2023. The survey will yield contiguous overlapping imaging of over half the sky in six 
optical bands ($ugrizy$, covering the wavelength range 320--1050 nm). 

\begin{figure}
    \centering
    \vspace{-2cm}
    \includegraphics[width=7cm]{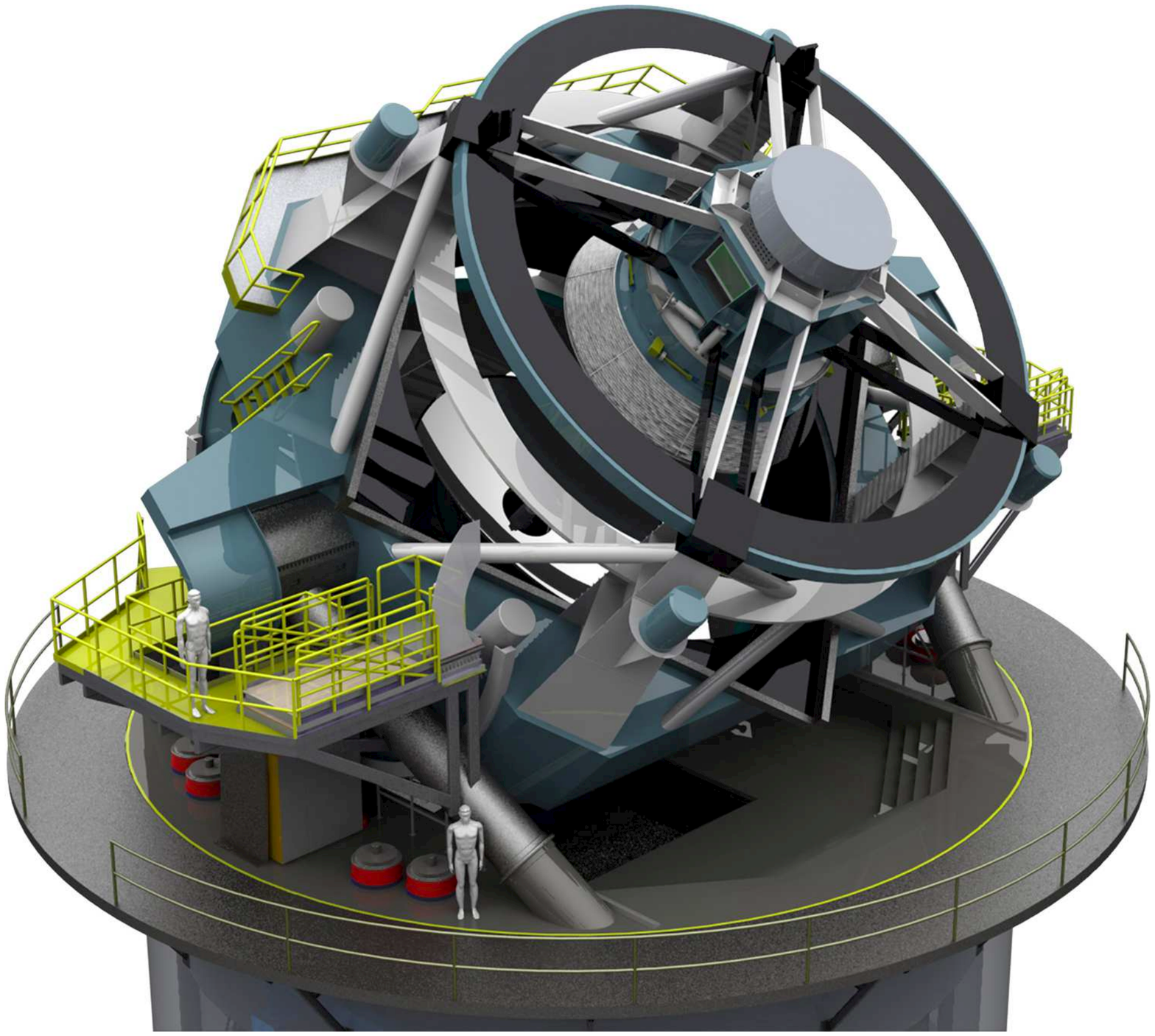}\\
    \vspace{-1cm}
    \includegraphics[width=14cm]{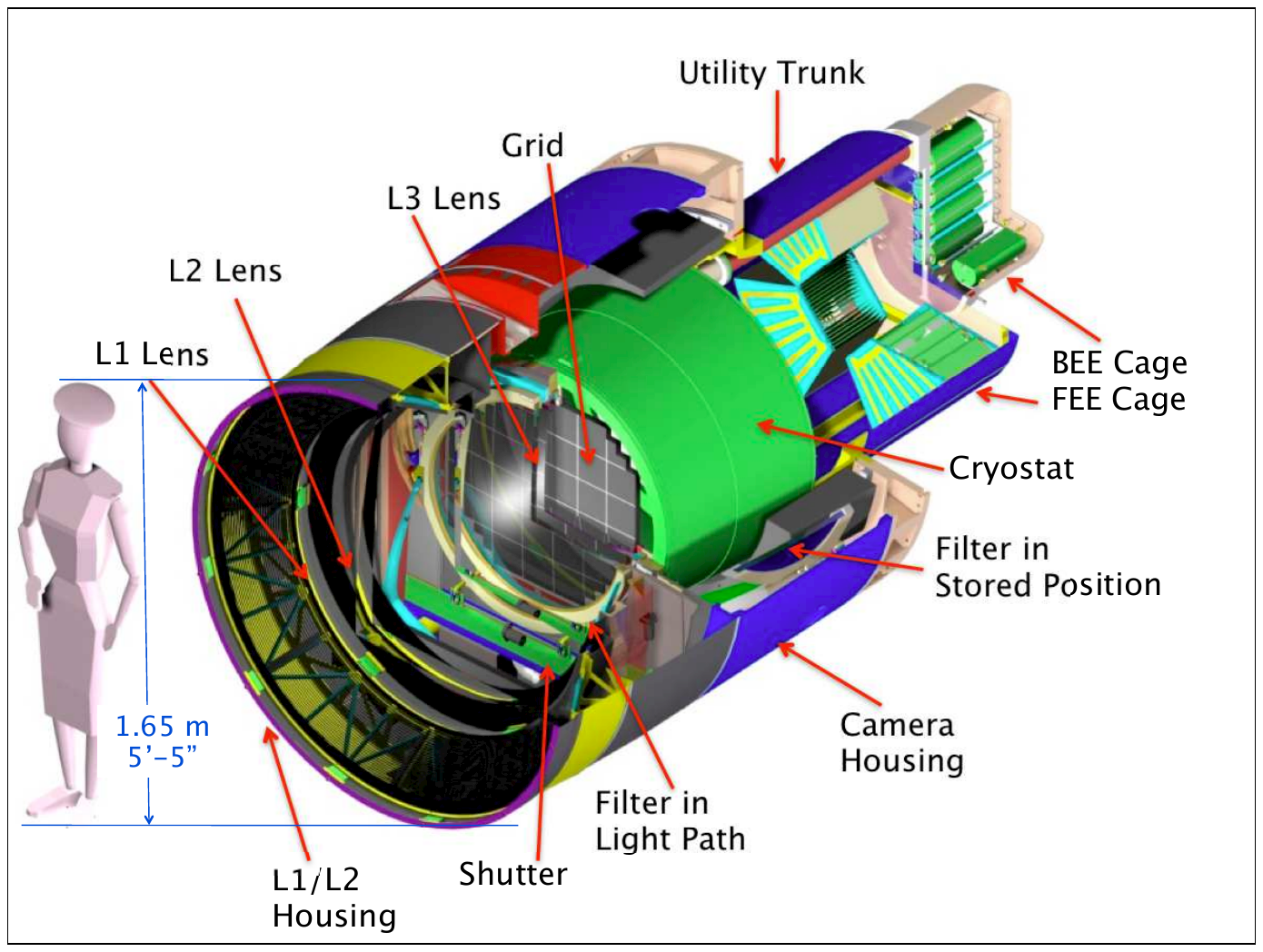}
\vspace{-5.5cm}
    \caption{Upper panel: baseline design for the LSST telescope. Lower panel: the LSST camera layout. Credit: LSST Corporation.}%
    \label{fig:lsst}%
\end{figure}

The LSST telescope uses a novel three-mirror design (modified Paul-Baker) with a very fast f/1.234 
beam (Figure \ref{fig:lsst}). The optical design has been optimised to yield a large field of view (9.6 deg$^2$), with 
seeing-limited image quality, across a wide wavelength band. Incident light is collected by the 
primary mirror, which is an annulus with an outer diameter of 8.4m and inner diameter of 5.0m 
(an effective diameter of 6.5m), then reflected to a 3.4m convex secondary, onto a 5m concave 
tertiary, and finally into three refractive lenses in a camera. This is achieved with an innovative 
approach that positions the tertiary mirror inside the primary mirror annulus ring, making it 
possible to fabricate the mirror pair from a single monolithic blank using borosilicate technology. 
The secondary is a thin meniscus mirror, fabricated from an ultra-low expansion material. All three 
mirrors will be actively supported to control wavefront distortions introduced by gravity and 
environmental stresses on the telescope. The telescope sits on a concrete pier within a carousel 
dome that is 30m in diameter. The dome has been designed to reduce dome seeing (local air 
turbulence that can distort images) and to maintain a uniform thermal environment over the course 
of the night. 

The LSST camera (Figure \ref{fig:lsst}) provides a 3.2 Gigapixel flat focal plane array, tiled by 189 4K$\times$4K CCD science 
sensors with 10 $\mu$m pixels. This pixel count is a direct consequence of sampling the 9.6 deg$^2$
field-of-view (0.64m diameter) with 0.2$\times$0.2 arcsec$^2$ pixels (Nyquist sampling in the best 
expected seeing of $\sim$0.4 arcsec). The sensors are deep depleted high resistivity silicon 
back-illuminated devices with a highly segmented architecture that enables the entire array to be 
read in 2 seconds. The detectors are grouped into 3$\times$3 rafts; each contains its own dedicated 
front-end and back-end electronics boards. The rafts are mounted on a silicon carbide grid inside a 
vacuum cryostat, with an intricate thermal control system that maintains the CCDs at an operating 
temperature of 180 K. The entrance window to the cryostat is the third of the three refractive lenses 
in the camera. The other two lenses are mounted in an optics structure at the front of the camera body, 
which also contains a mechanical shutter, and a carousel assembly that holds five large optical filters. 
The sixth optical filter can replace any of the five via a procedure accomplished during daylight hours.

The rapid cadence of the LSST observing program will produce an enormous volume of data ($\sim$15 TB of 
raw imaging data per night), leading to a total database over the ten years of operations of 50 PB for 
the raw uncompressed imaging data (100PB with processed versions), and 15PB for the final catalog database. The computing power required to process the data 
grows as the survey progresses, starting at $\sim$100 TFlops and increasing to $\sim$400 TFlops by the 
end of the survey. Processing such a large volume of data, automated data quality assessment, and archiving 
the results in useful form for a broad community of users are major challenges. The data management system 
is configured in three levels: an infrastructure layer consisting of the computing, storage, and networking 
hardware and system software; a middleware layer, which handles distributed processing, data access, user 
interface and system operations services; and an applications layer, which includes the data pipelines and 
products and the science data archives. 

The application layer is organised around the data products being produced. The nightly pipelines are based 
on image subtraction, and are designed to rapidly detect interesting transient events in the image stream 
and send out alerts to the community within 60 seconds of completing the image readout. The data release 
pipelines, in contrast, are intended to produce the most completely analysed data products of the survey, 
in particular those that measure very faint objects and cover long time scales. A new run will begin each 
year, processing the entire survey data set that is available to date. The data release pipelines consume 
most of the computing power of the data management system. The calibration products pipeline produces the 
wide variety of calibration data required by the other pipelines. All of these pipelines are designed 
to make efficient use of computer clusters with thousands of nodes. There will be computing facilities at 
the base facility in La Serena, at a central archive facility, and at multiple data access centres. The 
data will be transported over existing high-speed optical fibre links from South America to the USA. 

For a more detailed discussion, including optical design, the filter complement, the focal plane 
layout, and special science programs, please see the LSST overview paper \citep{LSSToverview}
and the LSST Science Book\footnote{Available from www.lsst.org/lsst/SciBook} \citep{LSSTSciBook}.

\subsection{Planned survey strategy and delivered data products} 

\noindent The LSST observing strategy is designed to maximise scientific throughput by minimising 
slew and other downtime and by making appropriate choices of the filter bands given the 
real-time weather conditions. The fundamental basis of the LSST concept is to scan the sky 
deep, wide, and fast, and to obtain a dataset that simultaneously satisfies the majority of 
the science goals. This concept, the so-called ``universal cadence'', will yield the main 
deep-wide-fast survey and use about 90\% of the observing time. The observing strategy for 
the main survey will be optimised for homogeneity of depth and number of visits. In 
times of good seeing and at low airmass, preference will be given to $r$-band and $i$-band 
observations. As often as possible, each field will be observed twice, with visits separated 
by 15-60 minutes. The ranking criteria also ensure that the visits to each field are widely 
distributed in position angle on the sky and rotation angle of the camera in order to minimise 
systematic effects in galaxy shape determination.

The current baseline design will allow about 10,000 deg$^2$ of sky to be covered using pairs of 
15-second exposures in two photometric bands every three nights on average, with typical 5$\sigma$ 
depth for point sources of $r\sim24.5$. For example, these individual visits will be about 
2 mag deeper than the SDSS data. The system will yield high image quality as well as excellent astrometric and photometric accuracy for a ground-based optical survey. The survey area will include 30,000 deg$^2$ with $\delta<+34.5^\circ$, with the 18,000 deg$^2$ main survey footprint visited over 800 times during 10 years. The 
coadded data within the main survey footprint will be 5 mag deeper than SDSS ($r\sim27.5$).
The main survey will result in databases including 20 billion galaxies and a similar number of 
stars, which will serve the majority of science programs. 
The remaining 10\% of observing time will be used to obtain improved coverage of parameter 
space such as very deep ($r \sim 26$) observations (e.g. optimised for SNe), observations with 
very short revisit times ($\sim$1 minute), and observations of ``special'' regions such as the 
Ecliptic, Galactic plane, and the Large and Small Magellanic Clouds. 

The LSST data system is being designed to enable as wide a range of science as possible. Standard 
data products, including calibrated images and catalogs of detected objects and their attributes, 
will be provided both for individual exposures and the deep incremental data coaddition. About 
20 billion objects will be routinely monitored for photometric and astrometric changes, and any 
transient events (non-recurrent objects with statistically significant photometric change; about 
10,000 per night on average) will be distributed in less than 60 seconds via web portals. For 
the ``static'' sky, there will be yearly database releases listing many attributes for billions 
of objects and will include other metadata (parameter error estimates, system data, seeing summary 
etc). 

LSST has been conceived as a public facility: the database that it will produce, and 
the associated object catalogs that are generated from that database, will be made 
available to the U.S. and Chilean scientific communities and public with no proprietary 
period. Negotiations are under way with prospective international partners to make 
LSST data more broadly available.

It is expected that the scientific community will produce a rich harvest of discoveries with LSST
data products. Many of the highest priority LSST science investigations will require organised teams 
of professionals working together to optimise science analyses and to assess the importance of 
systematic uncertainties on the derived results. To meet this need, eleven science collaborations 
have been established by the project in core science areas. As of the time of this contribution, 
there are over 500 participants in these collaborations, mostly from LSST member institutions.
Through the science collaborations, the astronomical and physics communities are
involved in the scientific planning of LSST deployment strategies.

\section{Overview of Synergies  }

\noindent The SKA and LSST are the two major ground-based survey telescopes of the next decade. They offer significant synergies,  in terms of both sky area and time-domain astrophysics, and are likely to be on the sky over much of the same time-period. The survey strategies for SKA1 and SKA2 are being developed in this book, but are likely to include $3\pi$ surveys over the same area as LSST, as well as thousands of square degrees to a greater depth than LSST. 

The LSST will leverage a large range of science from the SKA by providing approximate (photometric) redshifts for about 40 galaxies per square arcminute. Many of the existing SKA science applications assume that such photometric redshift information will be available. LSST is the only foreseen survey covering a large fraction of the southern hemisphere at an appropriate depth to provide this information. Euclid can further improve the photometric redshifts from LSST by supplying infra-red information. 

Conversely, the SKA can enable LSST to carry out additional exciting science. For instance, one of the challenges for LSST is to calibrate its photometric redshifts with precise redshift information. The SKA has the potential to provide a large amount of redshift information through observations of HI emission, which can then be used to calibrate LSST objects through cross-correlation \citep{newman, mcquinn}. SKA will also provide direct redshifts for training photo-zs for highly extinguished or featureless sources that fail to yield emission lines or spectral breaks in the optical/near-IR.

The synergy between the LSST and SKA will have a major impact for several disciplines. In cosmology, much effort is focussed on understanding the apparent accelerated expansion of the Universe. Forecasts currently often take the form of constraints on the dark energy equation of state; however in a decade we may be more concerned with testing the laws of gravity, or investigating new surprises yet to come. In any case, it will be necessary to confront theory with observations of the statistics of cosmological probes, to a large distance. The SKA and LSST are well-suited to each other to obtain the necessary observational cosmology data. 

In addition, cosmology in the next decade is likely to have reached the limit of what can be achieved simply by increasing the volume surveyed. Control of systematic errors will be paramount, and therefore cross-checks of quantities between the SKA and LSST will enable additional science. For best results, the data from both surveys can be cross-correlated to reduce systematic effects specific to one survey or the other.

When it comes to our attempts to understand how galaxies form, again we find that this quest is intimately tied to both optical and radio observations. The pivotal role of AGN in galaxy formation (see \citet{mcalpine}  and \citet{smolcic} in this book), and the presence of neutral hydrogen, can only be understood at radio wavelengths, whereas the star formation history can be extracted from both multi-wavelength optical studies and in an extinction-free way using radio continuum observations (see \citet{jarvis} in this book). LSST provides the necessary photometric redshift information that can be used to simultaneously extract approximate redshifts and star-formation rates. 

The SKA and LSST both have unusually high time resolution for survey instruments; both are likely to be able to process data in a matter of seconds. They also have a similar size of instantaneous field of view or beam. Therefore the potential for coordinated surveys to find transients opens up an exciting new regime in observational parameter space. 

%Many other areas will benefit from the synergy between SKA and LSST: the study of young and evolved stars, supernova remnants, and studies of the interstellar medium. Each chapter of this science book which refers to optical counterpart measurements is describing a potential application for combining SKA and LSST.

The SKA and LSST will both benefit greatly from other major facilities that will be observing the southern sky. Facilities such as Euclid and WFIRST will provide key near-infrared wavelength data coupled with high resolution that will aid both strong and weak lensing studies,  photometric redshift determination, and will allow estimates of the stellar mass of galaxies at $z>1$. The addition of eROSITA will provide a complementary approach to disentangling the AGN in radio continuum surveys. Complementary to the cosmological studies described here, the proposed all-sky ACTPol survey and future SPT surveys will provide high-resolution CMB polarisation and lensing maps along with a large increase in the number of SZ selected clusters. There are many powerful three-way synergies; here we will concentrate on the SKA-LSST axis.

\subsection{LSST-SKA Methodological Synergies} 

\noindent Although we focus predominantly on scientific synergies, the LSST and the
SKA will also present many methodological synergies.  Both experiments
will provide petabyte-scale observational data-sets recorded over
time.  Extracting all of the astrophysical information contained in
such big data-sets will be a considerable challenge.  Although the raw
observational data recorded by optical and radio interferometric
telescopes exhibit quite different properties, the underlying
techniques that will be used to analyse these data share many
similarities.  Bayesian analysis techniques are now of widespread use
in astrophysics \citep[e.g.][]{lewis:2002,feroz:multinest}; sampling
methods that scale to very high-dimensional settings, such as Gibbs
and Hamiltonian sampling \citep[e.g.][]{wandelt:2004, taylor:2008},
will be increasingly important in analysing big-data. Machine learning
techniques \citep[e.g.][]{ball:2010} will also play an increasingly
important role in tackling the curse of dimensionality that both LSST
and SKA data will suffer.  Supervised machine learning techniques can
be exploited to efficiently navigate these high-dimensional data-sets,
while unsupervised learning techniques can be used for dimensionality
reduction, allowing the data to effectively speak for themselves. 

The sparse structure of big data-sets can also be exploited.  Compressive
sensing \citep{candes:2006,donoho:2006} is a recent ground-breaking
development in information theory, going beyond the Shannon-Nyquist
sampling theorem by exploiting sparsity, and which has the potential
to revolutionise data acquisition in many fields \citep[for a brief
introduction see ][]{baraniuk:2007}.  Although the application
of compressive sensing techniques in astrophysics is not yet mature,
first applications for radio interferometric imaging have shown
considerable promise \citep[e.g.][]{carrillo}.  The effective
application of all of these analysis methodologies will be
instrumental in the extraction of scientific information from large
observational data-sets, such as those recorded by LSST and the SKA.
Both experiments will benefit greatly from sharing expertise,
analysis techniques and open-source numerical codes.

\section{Cosmology with LSST and SKA}

\noindent Here we will discuss the synergies we expect for major cosmological probes, in particular galaxy clustering (including intensity mapping), weak lensing and strong lensing. 

\subsection{Weak lensing}

\noindent LSST will have an unprecedented sample of 3 billion galaxies with high S/N and good colour-redshifts with which to generate dark matter tomographic maps. SKA1 can obtain 40 million galaxies for weak lensing, with SKA2 possibly matching LSST's number density \citep[see][in this book]{brown}. Overlapping optical and radio surveys such as those carried out by LSST, SKA1-MID and SKA2 have a
particularly useful synergy in terms of reducing and quantifying the
impact of systematic effects in weak gravitational lensing
analyses \citep{brown}. By cross-correlating the shapes of galaxies as
measured in the optical and radio surveys, one can eliminate
instrumental systematic effects that are not correlated between the
two telescopes \citep{patel}. Given the very different designs and modes of
operation of optical and radio telescopes, one would not expect their
instrumental systematic effects to be correlated, and so this offers a
route to measuring the cosmic shear signal in a very robust way.

In addition, radio surveys offer unique ways to measure the
lensing signal that are not available to optical telescopes. In
particular, both radio polarisation information and rotational
velocity measurements from HI observations can provide estimates of
the \emph{intrinsic} position angles of the lensing source
galaxies \citep{blain02, morales06, bb11}. Such measurements offer great potential to (i) reduce the
effects of galaxy ``shape noise'' (due to the intrinsic dispersion in
galaxy shapes) and (ii) to mitigate the contaminating signal from the
intrinsic alignments in galaxy orientations which is perhaps the most
worrisome astrophysical systematic effect facing future weak lensing
surveys. In addition to using this information in a combined analysis,
one could potentially use the SKA-based estimates of the intrinsic
alignment contamination to calibrate out the alignment signal in the
LSST lensing survey. 

Finally, the envisaged SKA1-MID and SKA2 surveys will probe a wider range of
redshifts than will be reached by LSST. They therefore provide
extra (high-redshift) tomographic slices with which the evolution of
structure at relatively early times can be probed. SKA can also push to 
high redshift by measuring the lensing distortion signal in HI
intensity mapping surveys \citep{pourt}. Thus, these high-redshift SKA lensing
experiments will naturally fill the gap between the traditional
optical lensing probes (where sources are typically located at $z\sim1$)
and the ultimate lensing source of the CMB at $z\sim1000$.

\subsection{Galaxy clustering }

\noindent The three approaches to SKA galaxy clustering benefit from LSST in distinct ways, which we will now describe.

\subsubsection{HI galaxy survey} 

\noindent
As discussed in the chapter on HI threshold surveys \citep{abdalla}, one approach to galaxy clustering with SKA is to measure redshifts from the HI line for a sample of individually detected galaxies. This provides us with a galaxy survey where the redshifts are known to very high precision (we assume a Gaussian error of $\delta_z = 0.0001$). There is a powerful synergy here between this spectroscopic-quality large scale structure (LSS) survey from SKA and the weak gravitational lensing (WGL) surveys from LSST (with or without the further improvements in systematics from combining with the SKA weak lensing survey, see section 3.1 above). The LSST WGL survey has poor redshift resolution (due both to photometric-quality redshifts and the inherently broad WGL geometric kernel) but has direct access to the true matter distribution in the Universe. In contrast the SKA LSS survey has very good redshift resolution but uses galaxies as biased tracers of the mass distribution. Combination of the two surveys allows us to control for uncertainties in galaxy bias and improve our knowledge of how mass clustering evolves with redshift compared to WGL alone. 

\begin{figure}
\includegraphics[width=1\textwidth]{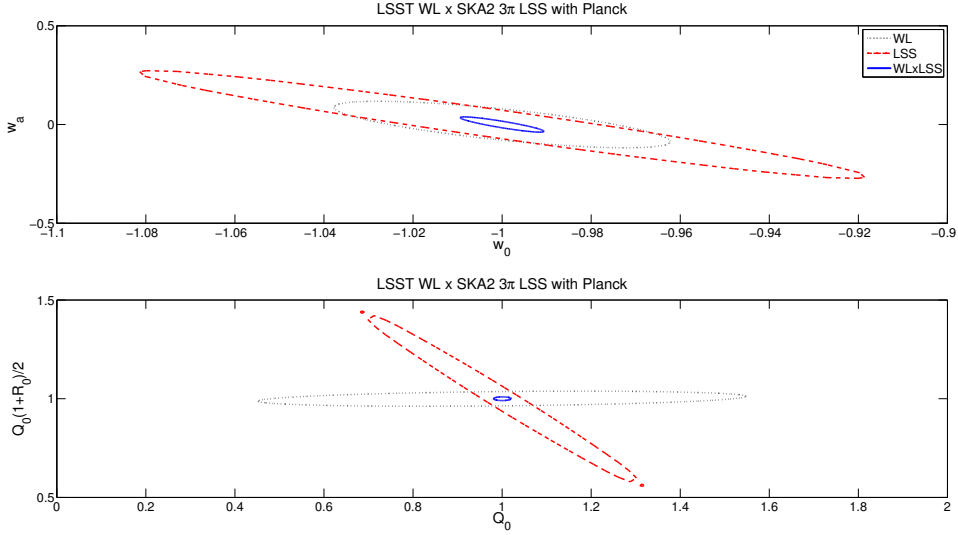} 
\caption{Constraints on dark energy ({\it top panel}) and gravity parameters ({\it bottom panel}) with SKA2 HI galaxy survey and LSST WGL survey, together with Planck priors.} 
\label{fig:wlxlss_ellipses}
\end{figure}

In Fig. \ref{fig:wlxlss_ellipses} we present results combining a specimen SKA LSS survey with an LSST WGL survey. Our WGL survey is assumed to have a source density of $n_g = 50$arcmin$^{-2}$ and to cover a redshift range of $0 < z < 3.0$. The galaxy redshift distribution is given by \citet{hojjati} and broken into 6 tomographic bins of approximately equal number density. The Gaussian scatter on the photo-z errors is $\delta_z = 0.03$.

We assume a LSS study with the full SKA2 survey over 30,000 deg$^2$. We have adopted the galaxy redshift distribution discussed in \citet{SKA:HI}; our forecasts assume 40 tomographic bins up to $z=2.0$.
We use the exact projected angular power spectrum, $C(\ell)$, formalism (not the Limber approximation) and include the effects of redshift space distortions (RSDs) according to the formalism of \citet{kaiser_1987}. Both these effects are neglected in the WL forecasts because the broad tomographic bins make their impact negligible. 
We use a maximum wavenumber of $\ell_{\rm max}= 3000$ for the WL analysis, and exclude non-linear scales in the LSS analysis using the cutoff $\ell_{\rm max}(z_{\rm med \; i}) = k_{\rm lin \; max}(z_{\rm med \; i}) \chi(z_{\rm med \; i})$. Here $z_{\rm med \; i} $ is the median redshift of tomographic bin $i$, $k_{\rm lin \; max}(z_{\rm med \; i}) = 0.132 z_{\rm med\; i} h^{-1}$ Mpc, and $\chi$ represents comoving distance.

Our Figure of Merit analysis forecasts constraints for a set of cosmological parameters: $\{ \Omega_{m}$, $\Omega_{b}$, $\Omega_{DE}$, $w_{0}$, $w_{a}$, $h$, $\sigma_{8}$, $n_{s}, b, Q_{0}, Q_{0}(1 + R_{0})/2 \}$. As well as the standard wCDM parameters, $b$ is a free amplitude on galaxy bias and $Q_{0}, Q_{0}(1 + R_{0})/2$ are parameterisations of deviations to General Relativity that modify the Poisson equation and the ratio of metric potentials; our ability to constrain these parameters quantifies our ability to test gravity on cosmic scales, see e.g. \citet{kirk}. When presenting constraints on dark energy we marginalise over the cosmological parameters and galaxy bias but keep the modified gravity parameters fixed. When presenting constraints on modified gravity we marginalise over cosmology, including $w_{0}$ and $w_{a}$, and galaxy bias. Priors consistent with the latest Planck temperature constraints are included.

It is clear from Fig. \ref{fig:wlxlss_ellipses} that the combination of SKA2 and LSST yields an impressive constraining power. These probes are very complementary; the WGL survey responds directly to the presence of matter, but has poor discrimination in redshift due both to the reliance on photometric redshift and the irreducible width of the lensing kernel along the line of sight. In contrast the LSS survey uses galaxies as biased tracers of the underlying dark matter distribution but has much greater resolution in redshift. The combined analysis makes use of the best features of both probes, gaining redshift discrimination from LSS while directly probing the growth of structure and the geometry of expansion through WGL.

Of course it is possible that systematic effects might affect each probe differently and lead to mis-aligned probability ellipses in the upper panel of Fig. \ref{fig:wlxlss_ellipses}, so careful tests and cross-checks for systematics are essential. On the other hand, another benefit of combined probe analysis is the ability to ``calibrate'' systematics of one probe using another. For example, galaxy Intrinsic Alignments (IAs) could be an important astrophysical contaminant of WGL measurements; but information from spectroscopic LSS surveys can be used to down-weight physically close galaxy pairs and mitigate the impact of IAs. 

The improvement over each probe alone is particularly pronounced in the modified gravity constraints; this  is the result of combining one probe sensitive to the bending of light (WL, sensitive to the sum of metric potentials $\Psi + \Phi$) and another probe using galaxies as non-relativistic tracers (LSS, sensitive to the Newtonian potential $\Psi$). This combination of sensitivities breaks a pronounced degeneracy in the MG parameter space and produces constraints far stronger than either probe alone. 

\subsubsection{Principal Component Analysis}

\noindent
The dark energy and gravity parameters constrained above are constant physical quantities as a function of time and scale. Instead, we can conduct the Principal Component Analysis (PCA) approach from the chapter by \citet{SKA:PCA} to examine how sensitive the LSST and SKA are to physical quantities. In particular, we can examine the sensitivity to the time evolution of the dark energy equation of state, $w(z)$, and the time and scale dependence of the effective Newton's constant $\mu(k,z)$, and the gravitational slip, $\gamma(k,z)$.

The details of our approach are given in \citet{SKA:PCA}. We consider several LSST cosmological probes: weak gravitational lensing, clustering, and the clustering-lensing cross-correlation; we use the number density and bias models given in \citet{hojjati}. In addition, we use the HI clustering surveys for SKA1-MID and SKA2, with number density, bias and survey parameters as in \citet{SKA:HI}, Tables 2 and 4). We make PCA analyses for each telescope independently in combination with Planck constraints, or in combination with each other and with Planck.

\begin{figure}
\includegraphics[width=1\textwidth]{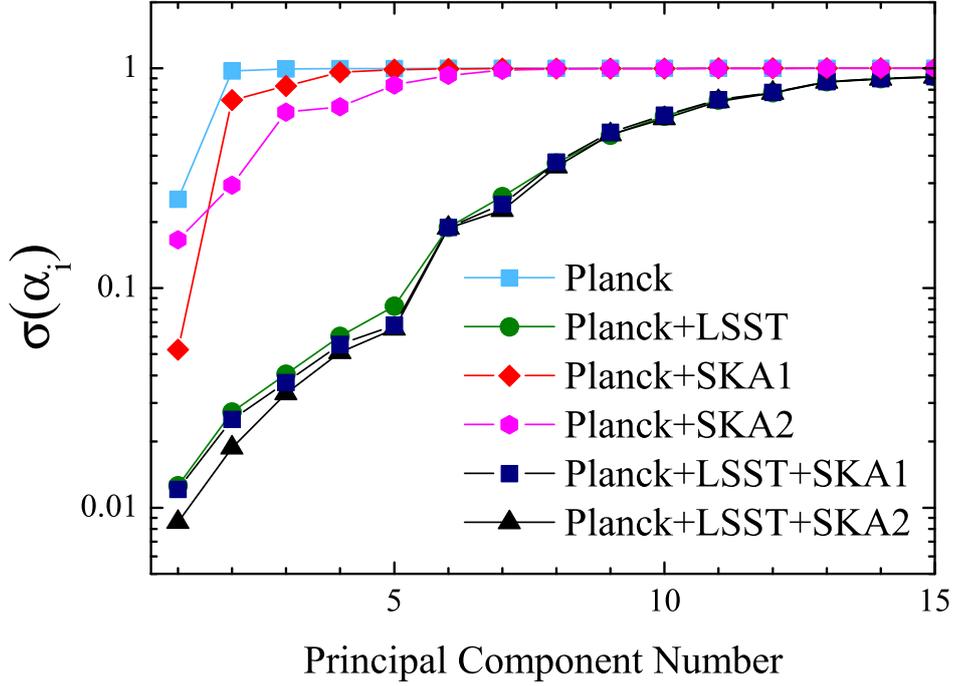} 
\caption{The forecast 68\% CL measurement error of the principal components, for the dark energy equation of state, for different survey combinations listed in the legend.} 
\label{fig:wpca}
\end{figure}

Firstly, we work in the context of General Relativity ($\mu=1, \gamma=1$) and calculate the Fisher matrix for each telescope combination, with  the cosmological parameters $\{ \Omega_{b} h^2, \Omega_{c} h^2, h, \tau, n_s, A_s, w_i \}$. After marginalising over the other parameters, we perform PCA on the $w_i$ bins. The results are shown in Figure \ref{fig:wpca}; we see that LSST+Planck is already excellent at constraining eigenmodes of the dark energy equation of state, with 5 modes constrained at the $\sigma<0.1$ level. SKA1 and SKA2 alone are not competitive for this purpose; however, SKA1 in combination with LSST substantially improves LSST constraints on the first few modes (by 70\% for SKA1 and 75\% for SKA2, for the best constrained mode).

\begin{figure}
\includegraphics[width=1\textwidth]{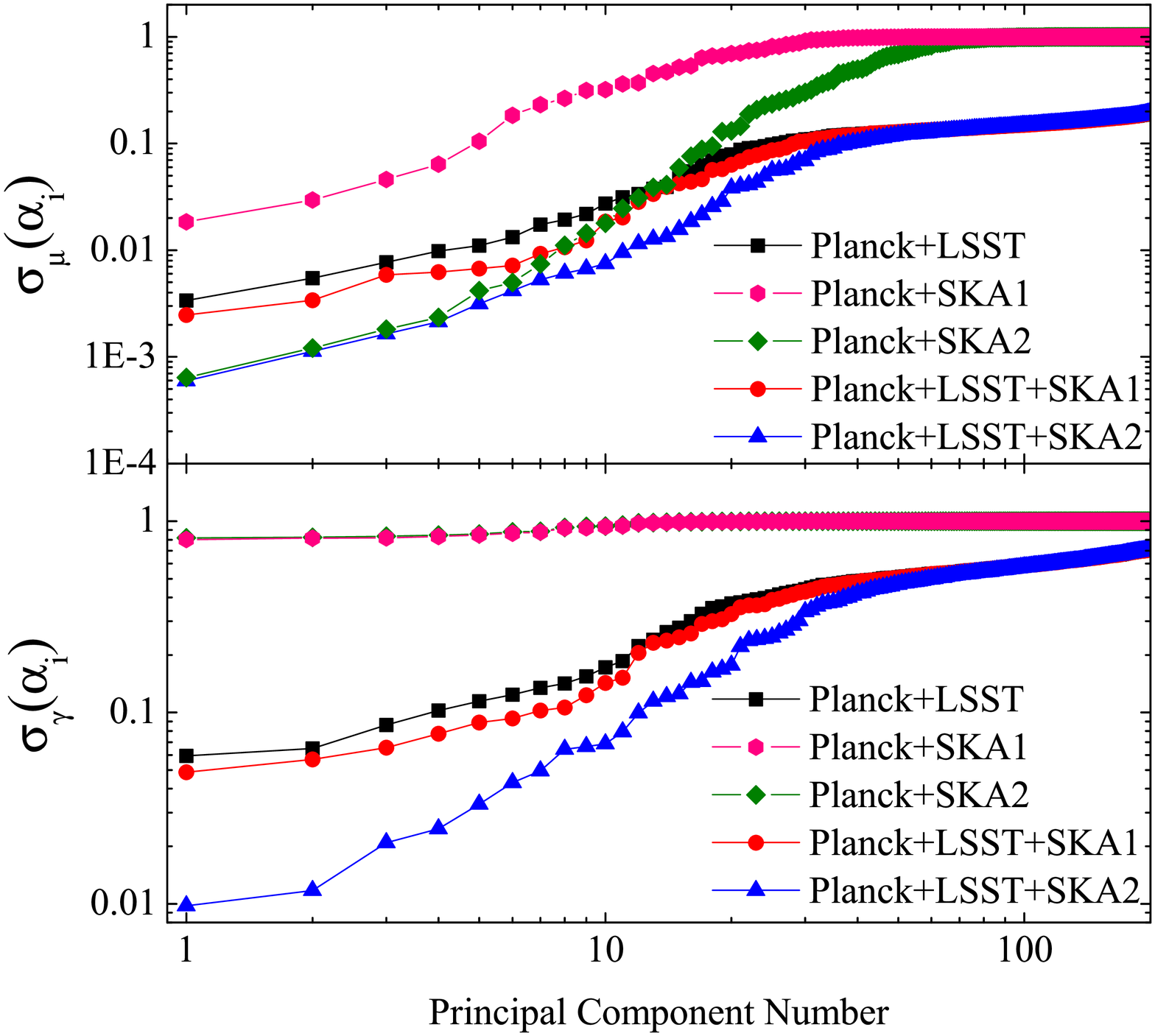} 
\caption{ The forecast 68\% CL error of the principal components of $\mu(k, z)$ (top panel) and $\gamma(k,z)$ (bottom panel), for different survey combinations listed in the legend.} 
\label{fig:mgpca}
\end{figure}

Next, we allow $\mu$ and $\gamma$ to vary as a function of time and scale, in addition to dark energy time variation. The results of the various PCA analyses, described in detail in \citet{SKA:PCA}, are shown in Figure \ref{fig:mgpca}; in this case, the synergy between LSST and SKA is very strong. 
While LSST+Planck alone can constrain 5 modes of the effective Newton constant $\mu$ at the $\sigma<0.01$ level, even LSST+SKA1+Planck can constrain 7, and LSST+SKA2+Planck can constrain 11 - with errors on the first mode reduced by a factor of 5. The situation is equally impressive for the gravitational slip $\gamma$; the addition of SKA1 to LSST+Planck reduces errors on the first mode by 20\%, and the addition of SKA2 reduces errors by a factor of 6. 

As in Section 3.2.1, the cause of this improvement is the different physical effects of $\mu$ and $\gamma$ on different cosmological probes. Weak lensing and the CMB are sensitive to combinations of the two metric potentials $\Phi$ and $\Psi$, which are affected by both $\mu$ and $\gamma$ modes. On the other hand, redshift space distortions measured in the SKA HI surveys are only sensitive to $\Psi$, and hence $\mu$. Thus we find that LSST+SKA are an excellent combination for future tests of gravity.

\subsubsection{Continuum survey}

\noindent As described in the chapter on cosmology with radio continuum surveys \citep{jarvis}, a new era for continuum cosmology is close to becoming a reality.  Among the numerous planned surveys are the LOFAR Surveys, the Evolutionary Map of the Universe (EMU), the MeerKAT-MIGHTEE survey and the Westerbork Observations of the Deep APERTIF Northern Sky (WODAN). These forthcoming experiments will provide us with a homogeneous all-sky continuum catalogue $>10$ times larger than the largest one hitherto available, and  SKA1 will be able to reach a factor an order of magnitude deeper over similar sky areas to these. 

However, radio continuum surveys do not provide any redshift information for the sources.  For cosmological purposes, this is a serious issue; to investigate cosmic acceleration, we require information about the time evolution of the Universe's expansion and structure growth.  For this purpose, \citet{camera1} proposed to cross-identify continuum radio sources with optical to near-infrared surveys (currently these include SkyMapper and SDSS). Even now such studies could be extended by incorporating data from other surveys, particularly at near/mid-infrared wavelengths where the VISTA Hemisphere Survey \citep{2013Msngr.154...35M}, 2MASS \citep{2MASS} and WISE \citep{WISE} can provide robust detections of low-redshift sources. LSST will further improve the prospects for cross-identification at higher redshift. By making these cross-matches, one can  separate the source distribution into a low- and a high-redshift sample, thus providing information on the evolution of cosmological parameters such as those related to dark energy.  This approach yields constraints more than four times tighter than in the case without redshift information. 

\begin{figure}
\center
\includegraphics[width=.8\textwidth]{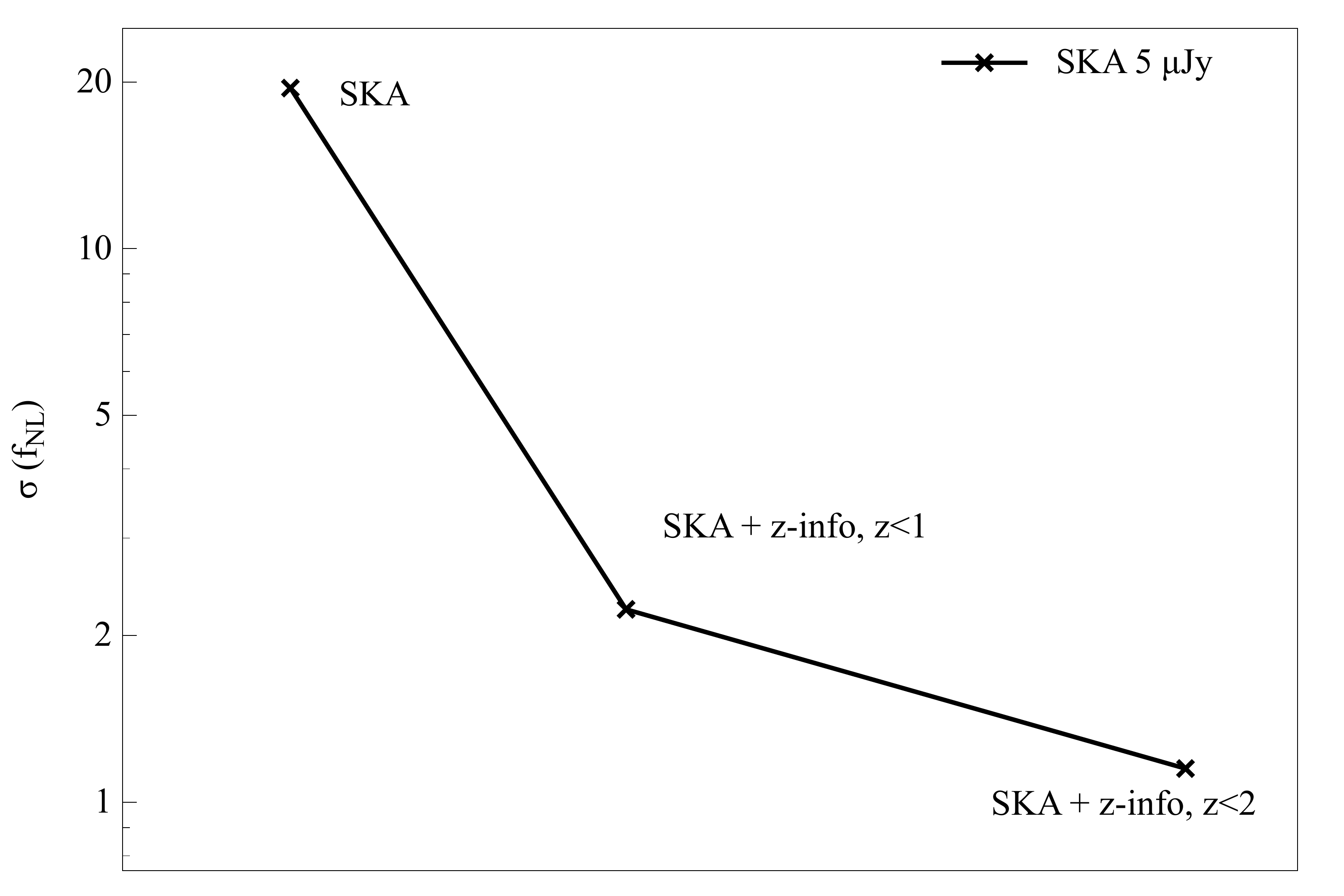} 
\caption{ Constraints on the non-Gaussianity parameter $f_{\rm NL}$ from the cross-correlation SKA-CMB, with and without redshift information provided by a photometric survey such as LSST. } 
\label{fig:isw}
\end{figure}

Using both the SKA and the LSST, we are also able to perform a cross-correlation of galaxy clustering  with the Cosmic Microwave Background, probing the integrated Sachs-Wolfe (ISW) effect. \citet{raccanelli} have provided promising forecasts for constraining the non-Gaussianity of primordial fluctuations using this probe with SKA and photometric surveys. This can be seen from Figure~\ref{fig:isw}, where we show how having redshift information will enable a tomographic ISW, which will help in pinning down the constraints on the non-Gaussianity parameter $f_{\rm NL}$.
We plot constraints for the SKA1 only case (no redshift information) for a 5$\mu$Jy rms survey of 30000 sq deg, and where redshift information is provided by a photometric survey such as LSST for SKA sources, up to $z=1$ and $z=2$. The error on $f_{\rm NL}$ is reduced from 20 to $\simeq$ 1, which compares favourably to the current current best constraint of $f_{\rm NL}=2.7\pm 5.8 (1\sigma)$ from Planck \citep{planck}.

\subsubsection{Intensity mapping} 

\noindent Intensity mapping of the redshifted neutral hydrogen 21cm emission line is an exciting new survey methodology for large scale structure. Instead of resolving many individual galaxies at high signal-to-noise, one instead uses low-resolution maps of the integrated emission from many unresolved galaxies to probe the large scales ($\sim 1^\circ$) corresponding to the baryon acoustic oscillations. Since the galaxy population hosting the HI emitting regions is a biased tracer of the underlying dark matter distribution, so too is the integrated HI signal, and redshift information is trivially obtained by the frequency of the emission. By making low-resolution maps over a substantial fraction of the sky, for many channels over a wide band, intensity mapping (IM) surveys are therefore able to rapidly reconstruct the large-scale redshift-space matter distribution over extremely large cosmological volumes, out to high redshift \citep{SKA:IM}.

While the IM methodology is not yet mature, a number of medium-size experiments that are either planned or in construction (e.g. CHIME, \citet{bandura}, BINGO, \cite{2012arXiv1209.1041B}) are expected to yield cosmological results substantially before SKA1 sees first light. With SKA1, a 10,000 hour, 25,000 deg$^2$ IM survey on either the MID or SUR arrays are projected to yield dark energy constraints that are competitive with a Dark Energy Task Force Stage IV galaxy redshift survey \citep{Bull:2014rha, SKA:BAO, SKA:RSD}. This could potentially be completed several years before LSST (and even Euclid), with a substantial overlap in survey area and redshift coverage.

While this would provide many of the same advantages as an overlapping galaxy survey in the radio, intensity mapping has a number of additional benefits in terms of synergies with LSST. Foremost is the significantly different set of systematic effects that one expects from making intensity maps rather than galaxy catalogues. A number of calibration and selection effects that are common to galaxy surveys, but not IM, can then be expected to drop out in cross-correlation, which will be especially powerful if, as expected, future large scale structure surveys are systematics-limited. IM surveys are also capable of spanning both lower and higher redshifts than LSST, and can potentially be used to `anchor' the LSST data by filling in some information in missing redshift bins -- for example, SKA1-MID can cover $0.4 \lesssim z \lesssim 3$ with Band 1, and $0 \lesssim z \lesssim 0.5$ with Band 2.

Because IM surveys probe a differently-biased population of galaxies to LSST, one can also benefit from multi-tracer analysis \citep{McDonald:2008sh, Abramo:2013awa, SKA:Camera}. For certain observables -- most notably redshift space distortions -- this allows the limits imposed by cosmic variance to be beaten, which is of particular importance for reconstructing the growth history of the Universe to high enough precision to test modifications to General Relativity \citep{SKA:RSD}.

\subsection{Strong lensing} 

\noindent The 2020s will be a new era for studying galaxy formation, the high redshift Universe and cosmology with strong gravitational lensing (see the chapter by \citet{SKA:strong} in this volume).  Currently, $\sim$500 strong lens systems are known, of which about 10\% are radio-loud. This sample will potentially increase with SKA and LSST to about 100000 in each waveband \citep[e.g.][]{marshall, oguri}. SKA alone will detect lensed sources in abundance, with a few per square degree accessible to SKA1 and many more to SKA2. These will include AGN and star forming regions via their continuum synchrotron emission, and gas clouds out to high redshift via their molecular line emission. The latter are particularly interesting for providing source redshifts and velocity fields.

Detailed multi-band optical and infrared imaging will be important both for finding and using this new sample of gravitational lenses. Having two surveys at very different wavebands reduces the false positive rate, which is likely to be the main problem with the next generation of lens surveys, because the radio-optical flux ratio of potential multiple lensed images will be a much better discriminant than anything available within a single waveband. (0.3"-resolution optical data would have drastically reduced the necessary followup in radio surveys such as CLASS (Myers et al. 2003), for instance). A secondary consideration is that detecting the lensing galaxy allows probabilistic arguments to be made about the likelihood of a lensing model for any surrounding objects, versus the hypothesis that the surrounding objects are non-lensed features such as star-forming regions. The deep multi-band photometry from the LSST will enable the lensing galaxies to be detected and their photometric redshifts to be measured in most of the SKA systems; lens galaxy redshifts and positions are needed for accurate modelling of their mass distributions. Once we have confirmed lens systems with good lens models, we will be able to test galaxy formation models, and explore source populations at high magnification. 

The synergy between the LSST and SKA is also important for the cosmological applications of strong lenses. Both telescopes will be capable of measuring independent gravitational time-delays from the variable optical and radio emitting regions, needed for the measurement of distances and so to test models for dark energy \citep[e.g.][]{suyu}. Multi-wavelength follow-up is very important in these studies, to characterise the properties of the lenses and their environment, which are needed to overcome systematics in the mass model. The radio monitoring is particularly powerful where the source is varying, since the lightcurves are not (or at least much less) affected by microlensing.

\section{Galaxy evolution with LSST and SKA}

\noindent With the onset of wide and deep field surveys across all wavebands, coupled
with high-resolution cosmological hydrodynamic simulations, much
progress has been made in understanding galaxy evolution over the past decade.
However, there are clear deficiencies in our understanding of galaxies
over the whole of cosmic time. At the bright end of the galaxy
luminosity function, we need to understand better the role of AGN feedback \citep{Fabian2012}, and how this may truncate or stimulate star formation in  the AGN host itself or in the surrounding
environment.
At the faint end, we need to understand the influence
of the environment, and how for instance hot and cold gas may get stripped from
galaxy haloes as they fall into the deep potential wells of clusters.

The combination of deep optical imaging from LSST, coupled with both
continuum and spectral line surveys with the SKA, offers a unique
opportunity to significantly impact on our understanding of the
formation and evolution of galaxies up to the highest
redshifts. Here we highlight some of the key
synergies that are also discussed at greater length in other chapters
in this volume.

\subsection{The evolution of activity in the Universe}

\noindent One of the key goals of the SKA continuum surveys is to provide a
complete census of the AGN and star-formation activity across cosmic
time. Radio continuum observations play a unique role for both of
these topics. For AGN, the radio traces highly energetic jets which
provide the mechanical feedback in many semi-analytic models and
hydrodynamic simulations of galaxy evolution. However, current radio surveys
do not provide a full picture of the impact of such sources. 
In particular, more depth is required to obtain information on the
detailed physics of distant radio sources, coupled with a full sampling of the
uv-plane to ensure that both large and small spatial scales are
sampled as required. On the other hand, the sources are 
rare and thus relatively wide surveys are also needed to ensure that
all environments at all redshifts are fully sampled.

Radio emission may provide
the most robust tracer of the star-formation rate in all galaxies.
This is because the radio waveband does not
suffer from dust extinction which is a limiting property of similar
studies using the visible waveband. Furthermore, the high-resolution,
sub-arcsecond imaging possible with the SKA ensures that AGN and star-formation activity can be  distinguished, and that radio sources
can be associated with optical and near-infrared counterpart
galaxies.

Therefore, continuum surveys with the SKA provide the necessary depth
and area to provide a complete census of AGN and star-formation
activity across cosmic time. However, what is missing from radio
observations is information on the redshift, stellar mass, and the many
other properties associated with the stellar populations within
galaxies (e.g. metallicity, age) which can be supplied by LSST. On the other hand, knowledge of the presence of an AGN from the SKA will ensure that AGN emission can be considered for modelling the SED at optical/near-infrared wavelengths.
Thus it is only by combining the SKA continuum surveys with
photometric redshifts and stellar properties from the LSST that we
will gain the fullest picture of galaxy evolution.

The LSST Deep Drilling Fields will provide extremely deep imaging data (AB$\sim 28$~mag) in four extragalactic fields, overlapping with deep near-infrared imaging from UltraVISTA \citep{McCracken2012} and VIDEO \citep{Jarvis2013} of LSST ($\sim 9.6$~deg$^2$ each). The LSST Deep Drilling Fields combined with the proposed deep SKA
continuum fields will allow us to trace Milky Way type galaxies up to
the highest redshifts with both facilities, sampling enough
cosmological volume not to be dominated by sample variance issues for this purpose. In particular, these LSST deep fields will be sufficiently deep to find galaxies at $z>6$, which will be ripe for CO redshift measurement with SKA Band 5. The
LSST all-sky survey coupled with the SKA all-sky survey will provide a
similar census of galaxy evolution at $z<1$, again sampling all of the
accessible volume.

\subsection{The evolution of hydrogen}

\noindent One of the unique aspects of the SKA is its ability to measure the
evolution of neutral hydrogen in the Universe. With SKA1 this can be
done in deep fields out to the limiting frequency for SKA-MID (350MHz
corresponds to a redshift of $z\sim 3$). In the wider area surveys the
ability to detect HI will probably be limited to around $z \sim
0.6$.

Together the SKA and LSST surveys could potentially provide the key information
on how quickly gas is turned into stars. For example,  in
combination with ALMA, the surveys could provide a continuous view of the path from
neutral (SKA) to
molecular gas (ALMA) through to star formation (radio continuum and LSST).
To be able to carry out such studies as a function of galaxy mass,
galaxy type and environment, will strongly enhance our understanding of the
evolution of galaxies.

\subsection{High-redshift galaxies and reionization}

\noindent
The LSST galaxy sample at $z\sim 5$ will provide an important calibration of stellar mass density and galaxy clustering, as a function of galaxy properties and environment. These can anchor interpretations of measurements of the brightness temperature fluctuations from the epoch of reionization with SKA. During the later stages of reionization, the brightness temperature is dominated by fluctuations in the neutral hydrogen fraction, which in turn depend on reionization-source properties and their clustering statistics \citep{Mellema:2012vz}. Measurements of such quantities during the epoch of reionization, $z=6-15$, with instruments such as VISTA, James Webb Space Telescope, and Hyper-SuprimeCam, will need such a lower-redshift normalization around the end of reionization to realize their full statistical potential. Deep observations with the James Webb Space Telescope to detect samples of the first galaxies will particularly complement the LSST sample by providing an anchoring point at the start of, or before, the epoch of reionization. The combination of the LSST, SKA, and James Webb Space Telescope will be key for joint constraints on galaxy evolution and reionization. 

The LSST Deep Drilling Fields will provide additional information during the epoch of reionization, $z>6$, in principle allowing studies of cross-correlation and co-evolution between galaxies and the brightness temperature/neutral hydrogen fraction from SKA. In combination with potential deep SKA continuum fields and complementary near-infrared data, the Deep Drilling Fields will allow the determination of key source population observables such as UV luminosity, star formation rate, escape fraction of ionizing photons, metallicity, and stellar mass (e.g. in combination with Spitzer or Euclid). Additional redshift measurements from SKA would complement this. 

As a baseline, by combining the source population data with the measured global brightness temperature signal from SKA, constraints can be placed on the fraction of  reionization that is provided by galaxies. If reionization occurs late, at $z\sim6-10$ as some evidence suggests, e.g.~\citet{2012MmSAI..83.1123P}, such that the brightness temperature fluctuations still trace the source statistics, cross-correlation of the LSST Deep Drilling Fields with SKA might be possible \citep{2009ApJ...690..252L, RPCWiersma:2012wi}. This would provide another way to constrain details of the reionization process, e.g. to what degree different galaxy types are responsible. Since LSST will characterise the source galaxy populations in great detail, the statistics of the brightness temperature field could also be directly correlated with galaxy properties. Such information is important for a full characterisation of the power spectrum of brightness temperature fluctuations, and hence for extracting the full cosmological information from SKA \citep{Mellema:2012vz}.

\section{Variables and transients}

\begin{figure}
\center
\includegraphics[width=1\textwidth]{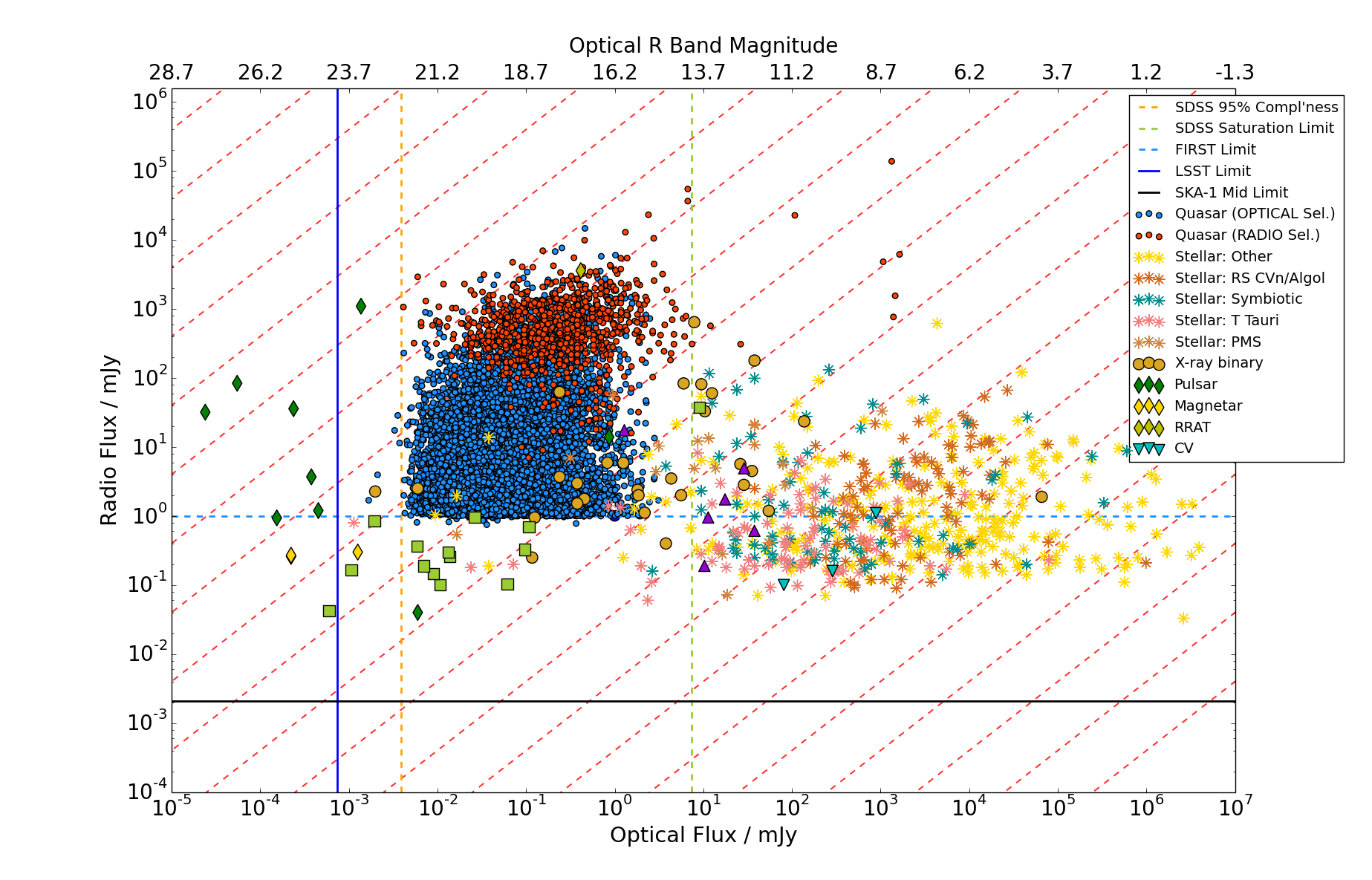} 
\caption{Radio vs optical flux density for a range of astrophysical variables, including a large set of SDSS AGN as well as stellar sources, pulsars and and active binaries. The sample is largely cut off by the limits of the SDSS and FIRST optical and radio surveys, respectively. Optical flux is immediately seen to be a good discriminator between the broad classes of source.
The solid mauve lines indicate the typical nightly sensitivities of LSST and SKA1-MID for a given field, demonstrating how much further into parameter space they will routinely push. For example, stellar sources will be detectable throughout the galaxy instead of just locally. Also indicated are the flux limits for a typical joint observation of a field with E-ELT and the dish component of SKA2 (albeit rather uncertain). From Stewart et al., in prep.} 
\label{radopt}
\end{figure}

\noindent One of the key science drivers for the LSST is the time variable Universe and transient astrophysics. In recent years this is an area which has also come to the fore in the key programmes being planned or implemented on the SKA and its pathfinders and precursors. LOFAR \citep{stappers}, MeerKAT, ASKAP and the MWA all have approved key programmes in the area of radio transients, both `fast' (coherent, found primarily in beamformed data) and `slow' (incoherent synchrotron, found mainly in image stacks). The chapters in this book by \citet{SKA:fender, SKA:macquart, SKA:corbel} provide an overview of the science and likely performance of the SKA transients surveys in both modes, with further, more detailed, case studies in the following transient chapters.
An earlier review of the prospects for radio transients with the SKA can be found in \citet{fender}.

However, it is clear that the value of finding radio transients is severely reduced if counterparts at other wavelengths are not also identified. Since the sources are by definition ephemeral, this usually means rapid identification, reporting and follow-up of radio transients (hence the strong push for automated commensal transient searching in \citet{SKA:fender}). Which wavelengths are best? In most cases, although X-ray and other counterparts will be important, the optical or infrared bands are most needed, as these can be readily compared against reference images and can identify candidates for spectroscopic follow-up (and possibly redshift measurement) in the most exciting cases. Probably the single highest priority, therefore, for radio transient follow-up, is to get an optical photometric measurement. Fig. \ref{radopt} presents a wide sample of the optical and radio fluxes, from sources ranging from stars to supermassive black holes, which are likely radio variables.

Fortunately, in the era of parallel wide-field SKA and LSST observations, the optical data will be readily available for most fields. For the kind of surveys being envisaged with SKA1, and the nightly sky sweeps of LSST, several fields per night will get much deeper than the parameter space explored in Fig. \ref{radopt} (mauve lines).

\section{Conclusions }

\noindent In this chapter we have examined the value of combining data and analyses with the SKA and LSST. We introduced the LSST, recognising it as one of the foremost survey telescopes of the next decade. We then discussed the synergies available between the SKA and LSST, at both the methodological and science result levels. 

In the field of cosmology, we discussed how weak gravitational lensing benefits from the combination of optical and radio shape measurements with radically different systematic effects present. We showed how the combination of LSST lensing and clustering, and SKA galaxy clustering with spectroscopic redshifts, provide improved constraints on dark energy and gravity parameters. We discussed how the synergy between LSST and SKA continuum and intensity mapping measurements also provides improvements on cosmological parameters. Strong lensing studies benefit from the combined ability of SKA and LSST to characterise both the lenses and sources in detail.

We have also discussed the benefits to studies of galaxy evolution, where SKA can provide information on redshift, neutral hydrogen, AGN and star formation, while LSST can provide complementary information about star formation, galaxy age and metallicity. LSST will provide an important calibration of galaxy properties for interpreting brightness temperature fluctuation data obtained with SKA from the epoch of reionization. The LSST Deep Drilling Fields in combination with SKA will help determine how galaxy populations reionize the Universe, thereby also providing information that will improve the SKA cosmology analysis. 

Finally, we have discussed the value of combining the LSST and SKA time domain in order to understand and discover a wide range of astrophysical variables and transients.

In conclusion, using both surveys gives:

\begin{itemize}
\item Complementary physical constraints (e.g. the sensitivity of LSST lensing to both metric potentials, and the SKA RSD measurements to the Newtonian potential alone);
\item Removal of systematics (e.g. cross-correlation of optical and radio lensing signals removing shape measurement systematics);
\item Cross checks of results; in the forthcoming period of concern about systematic effects, even using the two machines independently will provide important verification of the science results (c.f. the experiments at the Large Hadron Collider);
\item Mutual science support (e.g. LSST providing photometric redshifts for SKA continuum detections, and SKA calibration of LSST photo-zs via spectroscopic redshift cross-correlations);
\item A more complete picture (e.g. in galaxy evolution, LSST and SKA are sensitive to different components of the galaxy such as neutral hydrogen and stellar populations; in the time domain, the telescopes inform us of different aspects of the transient physics.) 
\end{itemize}

\section*{Acknowledgements}
\noindent DB is supported by the UK Science \& Technology Facilities Council (grant No. ST/K0090X/1). GBZ is supported by Strategic Priority Research Program "The Emergence of Cosmological Structures" of the Chinese Academy of Sciences, Grant No. XDB09000000, by the 1000 Young Talents program in China, and by the 973 Program grant No. 2013CB837900, NSFC grant No. 11261140641, and CAS grant No. KJZD-EW-T01. MS is supported by the South African SKA Project and the National Research Foundation. MLB is supported by a ERC Starting Grant (Grant no. 280127) and by a STFC Advanced/Halliday fellowship. AR is supported by the Templeton Foundation. Part of the research described in this paper was carried out at the Jet Propulsion Laboratory, California Institute of Technology, under a contract with the National Aeronautics and Space Administration.

\setlength{\bibsep}{0.0pt}
\bibliographystyle{apj}
\bibliography{SKALSST}

\end{document}